# The American Space Exploration Narrative from the Cold War through the Obama Administration[1]

Dora Holland[2,3] and Jack O. Burns[4,5]

[2] International Affairs Program, University of Colorado Boulder, Boulder, CO 80309

[3] Current mailing address: 11161 Briggs Court, Anchorage, AK, 99516.

[4] Center for Astrophysics & Space Astronomy, Department of Astrophysical & Planetary Sciences, University of Colorado Boulder, Boulder, CO 80309

[5] Corresponding author.

E-mail addresses: dora.holland@colorado.edu (D. Holland), jack.burns@colorado.edu (J. Burns)

**Abstract**

We document how the narrative and the policies of space exploration in the United States have changed from the Eisenhower through the Obama administrations. We first examine the history of U.S. space exploration and also assess three current conditions of the field of space exploration including: 1) the increasing role of the private sector, 2) the influence of global politics and specifically the emergence of China as a global space power, and 3) the focus on a human mission to Mars. In order to further understand the narrative of U.S. space exploration, we identify five rhetorical themes: competition, prestige, collaboration, leadership, and "a new paradigm." These themes are then utilized to analyze the content of forty documents over the course of space exploration history in the U.S. from eight U.S. presidential administrations. The historical narrative and content analysis together suggest that space exploration has developed from a discourse about a bipolar world comprised of the United States and the Soviet Union into a complicated field that encompasses many new players in the national to the industrial realms. The results also suggest that the United States was at a crossroads in 2017 on the issues of how it participates in this changing field of space exploration. From this, we make three observations regarding recent U.S. space policy: 1) there is a disconnect between stated policy goals in American space exploration efforts and the implementation of those goals, 2) the United States communicates mixed messages regarding its intent to be both the dominant leader in the field of space exploration and also committed as a participant in international collaboration, and 3) the United States cannot remain a true pioneer of space exploration if it does not embrace the realities of globalization and the changing dynamics within the field of space exploration. We conclude with three suggestions: 1) the U.S. government and NASA should critically examine space exploration priorities and commit to implementing a program that will further realistic and robust stated policy and goals, 2) the U.S. should re-examine its intention to play a dominant leadership role in space exploration and consider emphasizing a commitment toward active participation in international collaboration in space, and 3) the U.S should fully embrace the new paradigm of space exploration by lowering barriers like ITAR that hinder the competitiveness of the American space industry. The U.S. should review the challenges and advantages of collaborative endeavors with rising space-faring nations such as China and abandon Cold War era thinking, thus paving the way to Mars by encouraging the participation of many nations and space agencies on future human missions.

# 1. Background

In the summer of 2015, astrophysicist and science communicator Neil deGrasse Tyson conducted a podcast interview with the National Aeronautics and Space Administration (NASA) Administrator Charles Bolden. Tyson spoke of a shift in how the United States conducts itself in the arena of space exploration, explaining, "We have entered a new era...now we're in a different world, where other countries are rising in their presence on the space frontier." Bolden responded, "We [now] teach people, we act as a model for how people of different cultures, different races, different nationalities can in fact work together" [1]. In discussing NASA's recent emphasis on international collaboration as the model for how the United States (U.S.) will explore space in the future, Tyson and Bolden also highlight how this emphasis affects the meaning and purpose of space exploration for the Nation as a whole. They also contrast the past to the present, which suggests that the American space program is at an important crossroads.

The history of space exploration in the United States is often spoken of with a great sense of national pride. It is evident that the American persona, which values ideals such as innovation, curiosity, and determination, is reflected in how the U.S. has interpreted its role in international space exploration in the past and continues to be an important factor in the present. But how has the understanding of that role changed over time? By examining the rhetorical themes within U.S. national space policy, leadership statements, and policy recommendations throughout American history, we see that the way the United States presents its role in international space exploration directly mirrors the Nation's motivations in foreign policy and national security. The question this article explores is: how has the narrative of space exploration in the U.S. changed from the birth of the space program through the Obama Administration? By following the narrative of space exploration in the United States, it is possible to trace the country's development of national and international goals. As these goals have changed since the beginning of space exploration, so has the U.S. voice in space exploration. However, due to the struggle to actualize ambitious goals in space, the U.S. national space policy agenda often remains purely rhetorical. Today, America finds itself at a crossroads between the familiar position of geopolitical posturing to maintain national prestige and showing leadership in promoting the next steps toward a truly global, collective field of space research.

## 1.1 Space Exploration During the Cold War (1955-1991)

Although the inception of the American space program was influenced by the development of the space technologies of the Soviet Union (U.S.S.R.), it was not created to serve as a symbol of national pride or international competition. Under Eisenhower, U.S. space exploration was designed to represent American scientific achievements, not perpetuate political objectives [2]. Though Eisenhower's vision for American space exploration saw space as a scientific arena in which nations could further globally progress instead of competition, this vision would fall to the side as the relationship between space and nationalism became stronger in later administrations. On September 12, 1962, President John F. Kennedy proclaimed before a crowd of 35,000 that the United States had chosen to forge the pathway to the Moon [3]. He called out to the American pioneering spirit, stretching out a vision of success and prestige for the U.S. in the new frontier of space. The conquest of space was to become intrinsically tied to what it meant to be American, and with his rallying speech, Kennedy solidified the rhetorical link between U.S. prestige and dominance in space exploration [4]. This set the predominant tone for the Cold War era of U.S. space endeavors. Although interest in collaboration and cooperation remained, prestige and competition would take precedence. The political bipolarity of the Cold War, between the U.S. and the Soviet Union, was so prolific in every facet of society that space, the next step for human exploration, quickly became a proxy for competition.

*1.2 Space Exploration in the Post-Cold War Era (1991-2016)*

The aftermath of the success of the *Apollo* missions left the United States as the self-perceived "winners" of the Space Race, but rather than serving to encourage the American space program to continue making monumental achievements in space, *Apollo* and the "lavishness" of the Kennedy era became a benchmark to hold up as a comparison to the relative frugality of the modern era [5]. The United States has moved into the next stage of American space exploration, in which NASA stretches to achieve similar accomplishments to the lunar landings, but with an inadequate budget and decreasing public interest. In the post-Cold War era, the United States struggles to decipher its identity as a space-faring nation [6]. It hesitates to relinquish its status as the preeminent leader, but yet is also unwilling to commit to space on the same level as the *Apollo* period. The new paradigm of space exploration that has arisen in the post-Cold War period also influences the decision making of the U.S., complicating the question: what does the United States want from space exploration?

## 2. Methodology

In order to investigate how the narrative of space exploration has changed over time, we conducted an analysis of forty documents and looked at the policies and rhetoric concerning U.S. space exploration across eleven U.S. presidential administrations, from the Cold War era through the Obama Administration. We specifically examined three types of documents: policy releases from various presidential administrations, statements made by those presidents and other prominent voices in space exploration leadership, and policy recommendations made to the U.S. government for advised action in space exploration.

To more rigorously evaluate how the rhetoric and priorities of space exploration in the United States have changed throughout the course of space history in the United States, we conducted a summative content analysis on the documents using five general themes (Table 1). This method was chosen in order to analyze space policy documentation over a long period of time, because it is a useful tool for identifying how different patterns emerge from a larger body of material [7]. Content analysis is used when examining how language embodies intent, attitudes, and biases beyond its literal textual meaning. Hsieh and Shannon write that a summative content analysis is the analysis of overarching patterns within the content of a text or speech [7]. This type of analysis can be used to understand the larger picture, rather than focusing solely on isolated words and sentences, allowing the researcher to understand the latent meaning of the content as a whole [7]. By examining the presence or absence of certain themes within these documents, historical changes emerged that can be interpreted as an evaluation of policy rhetoric surrounding the role of space exploration in the U.S. over the course of time.

Five themes were selected for coding after allowing rhetorical patterns to emerge from a close reading of the collected data. Following Hsieh and Shannon's example, the method of content analysis aided in evaluating the rhetorical patterns that emerged. In addition to revealing the evolution of these rhetorical themes, the method of content analysis aided in the understanding of the nuanced nature of space exploration history. Each document was coded for those themes directly, and it was recorded how many times each theme occurred within a document. Those numbers then added up to total counts per theme per administration.

For the purposes of this study, prestige and leadership were understood to be two separate themes. Prestige was identified as rhetoric in which the United States utilized space exploration to increase its status globally, and leadership were statements in which the United States expressed specifically a desire to have a leadership role in space exploration.

The summative content analysis in this study is from the Eisenhower through Obama Administrations. We acknowledge the current administration but found that it would be premature to try to apply our analysis to the Trump administration after only one year.

*Table of Codes*

| Theme | Examples Illustrative of Coded Theme |
|---|---|
| **1:** Competition with the Soviet Union | <ul><li>The United States is in direct competition with the Soviet Union in the field of space exploration.</li><li>The Soviet Union, in developing its own space technologies, influences the decision making of the United States in its respective space endeavors.</li><li>The United States needs to "beat" the Soviet Union in space exploration</li><li>E.g. "…the Soviets are ahead of the United States in world prestige attained through impressive technological accomplishments in space." [8]</li></ul> |
| **2:** American Prestige | <ul><li>The actions of the United States in the field of space exploration are directly tied to the American persona</li><li>American nationalism is tied to accomplishments in space exploration</li><li>Achievements in space exploration are a source of national prestige</li><li>It is important to achieve successes in space exploration for the psychological benefits, and in order to increase prestige</li><li>E.g. "Considerable prestige and psychological benefits will accrue to the nation which first is successful in launching a satellite." [9]</li></ul> |
| **3:** International Collaboration | <ul><li>There is collaboration or cooperation with other international space agencies on the part of the United States.</li><li>There is a goal of partnering with international collaborators on space-related projects.</li><li>International collaboration is important to U.S. interests in space exploration.</li><li>E.g. "We should encourage greater international cooperation in space." [10]</li></ul> |
| **4:** American Leadership | <ul><li>The U.S. is and/or has been a leader in the space field.</li><li>There is a goal of maintaining American leadership in space exploration.</li><li>Space will be a peaceful arena specifically with American leadership.</li><li>The United States will continue paving the way for space exploration globally.</li><li>E.g. "…to become the world's leading space-faring nation." [3]</li></ul> |
| **5:** A New Paradigm | <ul><li>There is the sense of a new era of space exploration.</li><li>The space exploration field is a multi-national arena that encompasses many players, rather than the traditional bipolar relationship between the U.S. and Russia.</li><li>Space exploration is a changing field that does not resemble the past.</li><li>E.g. "The challenges facing our space program are different, and our imperatives for this program are different than in decades past." [11]</li></ul> |

Table 1. Five themes were used for the content analysis of U.S. policy releases, leadership statements, and policy recommendations related to space exploration within eight administrations from the Cold War to the Obama Administration.

## 3. Results

Using the five themes given in Table 1, we found that these themes changed significantly from the Eisenhower to Obama administrations. The raw counts of each theme, per each of the eight administrations, from the content analysis are shown in Table 2 are depicted as a histogram in Figure 1.

*Counts of Thematic References in Analyzed Documents for Each Administration*

| | **Theme 1** Competition with the Soviet Union | **Theme 2** American Prestige | **Theme 3** International Collaboration | **Theme 4** American Leadership | **Theme 5** A New Paradigm |
|---|---|---|---|---|---|
| **Eisenhower Administration** | 21 | 11 | 12 | 1 | 0 |
| **Kennedy Administration** | 10 | 4 | 4 | 10 | 0 |
| **Johnson Administration** | 5 | 3 | 6 | 4 | 0 |
| **Nixon Administration** | 10 | 7 | 12 | 6 | 0 |
| **Ford Administration** | 1 | 3 | 5 | 2 | 1 |
| **Carter Administration** | 5 | 4 | 6 | 4 | 2 |
| **Reagan Administration** | 3 | 6 | 26 | 23 | 6 |
| **George H. W. Bush Administration** | 0 | 11 | 18 | 16 | 17 |
| **Clinton Administration** | 0 | 3 | 15 | 5 | 3 |
| **George W. Bush Administration** | 0 | 4 | 14 | 2 | 5 |
| **Obama Administration** | 0 | 4 | 12 | 16 | 11 |

Table 2. Content analysis results depicted as a table of the raw counts for each identified theme, showing the number of references in each of the five themes within the categories of policy documents, policy recommendations, and leadership statements for eight U.S. presidencies.

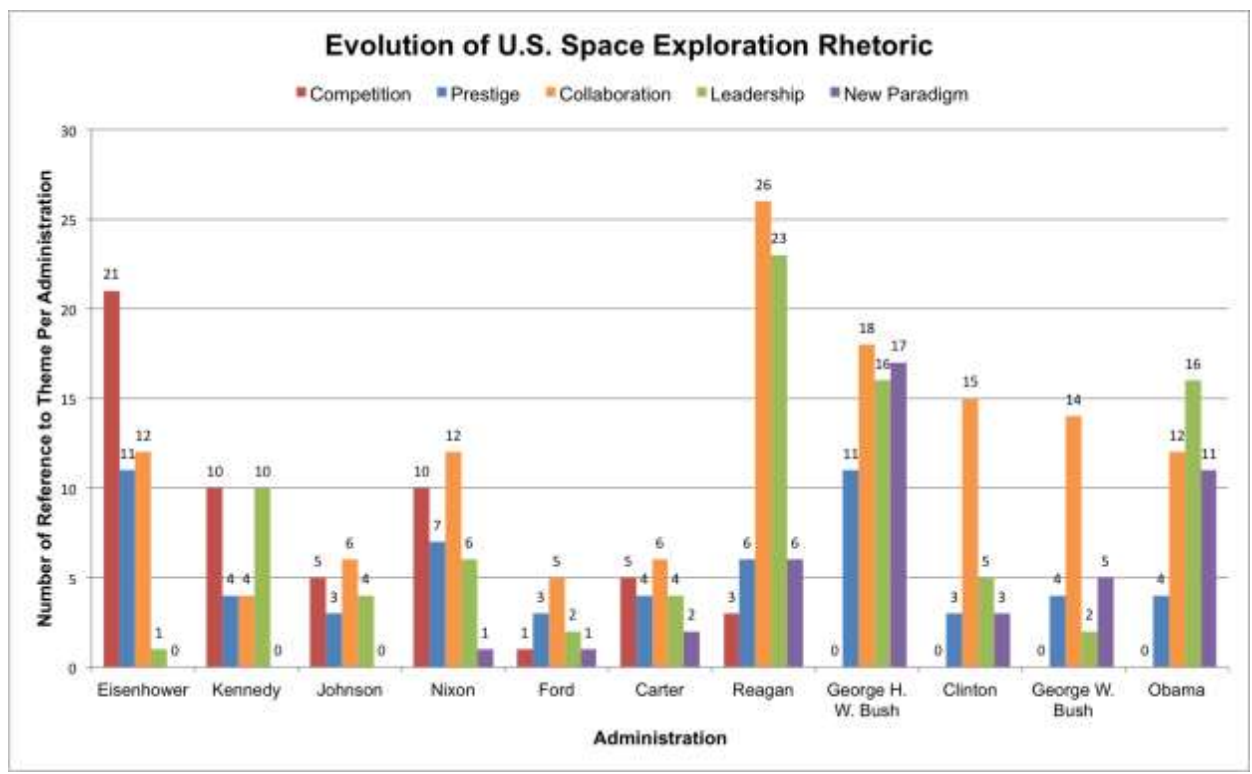

Figure 1. Summative content analysis results depicted as a histogram, showing the evolution of the identified themes throughout each of the eight administrations. The histogram demonstrates trends for certain themes, such as a decrease in competition over time and an increase in a new paradigm of space exploration. Other themes, such as prestige, collaboration, and leadership, are more complicated and do not show obvious trends.

While some simple trends are apparent in the data, such as the dominance of competition as a theme in early U.S. space program rhetoric and the emergence of the theme of a new paradigm in later years, the results show that space exploration rhetoric is complex and varies from one administration to another. Each new U.S. president sets out their own goals for space exploration, and there is variation and, in some cases, conflicting messages. Some administrations may desire to increase American prestige and leadership in space exploration, while others seek to reach out to other space-faring nations and develop collaborative ties. The analysis of these themes reveals how space exploration is an arena where many different, sometimes mutually exclusive, ideals and values interact and conflict with each other. Understanding the development of these themes is important to comprehending the development of the symbol of space exploration in the United States as a whole.

### 3.1 Theme 1: Competition with the Soviet Union

The space exploration narrative begins with a strong presence of Theme 1 in the Eisenhower through Nixon administrations. The psychological impact of the Soviet Union's launch of *Sputnik 1* is evident in the Eisenhower era, and despite President Eisenhower's dismissal of the *Sputnik* successes as anything meaningful, competition with the Soviet Union dominated early American efforts in space exploration. Well before the launch of the world's first space satellite, the National Security Council brought recent Soviet activities in developing space technology to Eisenhower's attention and explained that future explorations into space would be a race against the Soviet Union that the United States could not afford to lose [9]. After Eisenhower, the Kennedy administration fully embraced the Space Race, and set out to specifically match or exceed the Soviets' space program [12]. In the Nixon era, Congress requested an entire study titled, "United States and Soviet Rivalry in Space: Who Is Ahead, and How Do the Contenders Compare," [13] in order to determine who had the upper hand in space exploration. In the earlier days of U.S. space efforts, space exploration served as a competitive proxy between the United States and the Soviet Union, symbolizing the general rivalry that was the Cold War. The presidents and policymakers of these eras were unable to escape the overall atmosphere of "beat the Soviets," and even if the Soviet Union was not explicitly mentioned when space exploration goals were stated, the Cold War and competition with the Soviet Union was heavily embedded in the American psyche. However, Theme 1 decreases over time, until it disappears entirely after the Reagan administration. This signifies the shift in space exploration rhetoric from competition to collaboration, as the world moves from the Cold War to the post-Cold War era. From our document analysis, we can surmise that competition is no longer the primarily dominant theme in space exploration and other rhetorical themes have risen to make the overall field more complicated.

*3.2 Theme 2: American Prestige*

The consistent presence of Theme 2 in space exploration rhetoric directly demonstrates how space exploration serves a metaphorical purpose in the United States. Space exploration successes are often used as a symbol of the strength of American ideals and values, and those successes bring prestige to America. During the Eisenhower era, the United States was predominantly reacting to recent Soviet Union advances in space technology - specifically the launch of *Sputnik 1*. While Eisenhower himself devoted little attention to this event, his administration was well aware of the psychological benefits of being the first nation to launch a satellite [9] and related this back to the need of the U.S. to improve its own international image with successes in space. Before the launch of *Sputnik 1*, as both the United States and the Soviet Union were making preparations for launching satellites during the International Geophysical Year from 1957-1958, both countries understood that the first nation to launch would achieve great psychological and prestige benefits. After the Soviet Union achieved those benefits, the National Security Council acknowledged that the Soviet Union had surpassed the United States in terms of prestige and advised that the United States should work to improve its own image as a successful space-faring nation [14]. After a handful of presidential administrations and impressive U.S. space missions like *Apollo*, the George H. W. Bush administration had the ability to look back on 30 years of history in space exploration in the United States with praise. The President was fully aware that space exploration served as a symbol of American prestige and was proud of America's space exploration program, which he acknowledged in his speech on the 20th anniversary of the *Apollo 11* moon landing [15]. However, at this point in time, the United States was beginning to look to the future in order to plan ahead but maintaining American prestige in space was revealed to be an important goal during the George H. W. Bush administration. Our document analysis suggests that prestige remains an important theme throughout U.S. space exploration history, and although it is never at the forefront of rhetoric, it plays an important role in the decision making of the United States in its goals for its space program.

*3.3 Theme 3: International Collaboration*

International collaboration as a theme was continually present in U.S. space policy, even during the earlier days of competition with the Soviet Union. It also dominated rhetoric during the Cold War era administrations of Nixon and Reagan. The earlier rhetoric of international collaboration was generally limited to calls for collaboration with countries "friendly" to the United States. However, there were occasional dealings with the Soviet Union for exchange of some minimal technical data [16] within the Kennedy administration. Under Nixon and Reagan, Theme 3 becomes the key theme in space exploration rhetoric. As the Cold War reaches its height, America looked to its allies to maintain momentum against the Soviet Union in space endeavors. Nixon himself stated he believed that the successes of space exploration "should be shared by all peoples" [17], and Reagan reiterated this sentiment during his 1984 State of the Union Address when he stated, "NASA will invite other countries to participate so we can strengthen peace, build prosperity, and expand freedom for all who share our goals" [18]. The most important American goal during this period of history was without a doubt to achieve advantage over the Soviet Union. However, once the Soviet Union fell and began to look less like an enemy and more like a possible partner, Russia was included in this list of possible partnerships for milestone space exploration endeavors like the Space Station Freedom (or as it is later to be known, the International Space Station). Theme 3 maintains its importance over time as globalization and the increasing number of space-faring nations necessitates international collaboration. Additionally, due to the multiple interpretations of what it means to collaborate internationally, this theme is constant throughout all administrations. This theme also dominates rhetoric by a wide margin during the Clinton and George W. Bush administrations, where the construction and assembly of the International Space Station is paramount. By the post-Cold War era, the emphasis has shifted away from projects that America can achieve solely on its own, toward more collaborative endeavors.

*3.4 Theme 4: American Leadership*

In the documents we found that American leadership in space is also consistently present in U.S. space exploration rhetoric. In the Kennedy administration, it ranks evenly with competition as the dominating theme, demonstrating that not only did the U.S. view space exploration as a struggle with the Soviet Union, but that there was also the goal of becoming the leader of the field. This theme is relevant throughout the rest of the Cold War and reaches a peak during the Reagan Administration. However, instead of decreasing as the number of space-faring nations increases, and themes like international collaboration become important, Theme 4 maintains its prevalence. The crests in this theme during the Reagan and George H. W. Bush administrations are due to the fact that the United States was enjoying its established leadership role in space exploration and sought to maintain this role and exert dominance over the Soviet Union. Reagan remarked that "being a leader in space is a very wonderful accomplishment" [18], and the National Space Council under the George H. W. Bush administration advocated that the "fundamental objective" of U.S. efforts in space exploration is leadership [19]. Theme 4 decreases somewhat during the Clinton and George W. Bush administrations but re-emerges during the Obama administration. This underscores that although the United States advocates for international collaboration in space, it still seeks to be a leader in the space exploration and commercialization.

*3.5 Theme 5: A New Paradigm*

A "new paradigm," as a theme, is entirely absent during the Cold War era, as not only was this a period in which the status quo is competition between the U.S. and the U.S.S.R., but the United States did not yet have the history to draw comparison between the past and the present. It first appears as an important theme during the Reagan administration, at the tail end of the Cold War, when the National Commission on Space explained that power in space was rapidly proliferating as more and more nations made significant advances in space exploration with their own space capabilities [20]. However, Theme 5 materializes as early as the Nixon administration, as the U.S. was descending from the height of the *Apollo* missions, and wondering where to go next. Theme 5 reaches its peak under the George H. W. Bush administration, directly after the fall of the Soviet Union, when the world was in the post-Cold War shock. As the United States grappled with its direction for the future in order to achieve its space exploration goals, it acknowledged that this was no longer as simple as surpassing one identified rival. Theme 5 stands out in the rhetoric of the Obama administration. The increasing importance of private industry in space exploration adds to the complexity of space exploration as a field. Globalization not only complicates the arena with the inclusion of more players, but also competing interests in the industrial side. Under the Obama administration, the president acknowledged the multipolar nature of modern space exploration, as well as the significant role of private industry in this field. There is a direct comparison to the past in our current era, where our challenges are now different [21], and 50 years after the establishment of NASA, the space exploration narrative has evolved significantly [22].

**4. Discussion**

When examining the patterns that emerge from the five themes of the content analysis, it is important to consider how these themes interact with each other. Understanding this relationship allows us to develop a larger picture of the evolution of space exploration as a whole, with these different moving parts representing shifts in the interests and goals of the U.S. over time. By looking at what the United States desired to achieve recently in space exploration in contrast to its stated goals in the past, we can understand and make suggestions for how the country should move forward in the future.

Competition becomes less present in U.S. space exploration rhetoric as other themes, such as international collaboration, American leadership, and a new paradigm become more prevalent. It is plausible that the fall of the Soviet Union and American interest in partnering with the new Russia contributed to this shift in U.S. goals for space exploration, and later on, this interaction between competition and other themes could be the result of the field of space exploration becoming more complex and diverse.

Another theme that recedes over time is the theme of prestige, which is an important tool used by earlier leaders to gain the support of the American public by promoting American values through space exploration. Although the United States wishes to remain a leader in space exploration as time goes on, prestige as a theme falls to the wayside in favor of international collaboration. The rhetoric shifts from simply flaunting the national image to enthusiasm for Americans as helping others grow their own space capabilities. This transition is also related to the theme of a new paradigm. As many more interested parties enter into the space exploration arena, the United States made the decision to de-emphasize nationalism in favor of improving international relations and promoting leadership qualities.

When competition and prestige disappear as space exploration themes, and international collaboration and a new paradigm rise, one might expect that references to American leadership decreases as well. If leadership is seen as taking a commanding role it logically would be in conflict thematically with collaboration. The relationship between leadership and collaboration themes is discussed later in this section. Curiously, under the Obama administration the theme of leadership rose once again. This may reflect an administration attempting to navigate an increasingly complex arena of space exploration or may be indicative of Obama's values and goals for his legacy. Whatever the cause, the documents analyzed regarding Obama's 2010 National Space Policy mention leadership as a goal six times, communicating the intention to maintain U.S. dominance in the field of space exploration in the future.

The analysis and interactions among themes point to three observations regarding U.S. space policy through the Obama Administration. These observations are: 1) there is a disconnect between stated policy goals in American space exploration efforts and the implementation of those goals, 2) the United States communicates mixed messages regarding its intent to be both the dominant leader in the field of space exploration and committed as a participant in international collaboration, and 3) the United States cannot remain a true pioneer of space exploration if it does not embrace the realities of globalization and the changing dynamics within the field of space exploration.

*4.1 Observation 1: Disconnect Between Policy and Implementation*

After the successes of the *Apollo* era, the United States has had difficulty in clearly defining its goals in space exploration. As space policy has become less of a priority relative to other national interests, commitment and follow-through with stated policy goals is often not as consistent. Whereas in the past space was more of a national priority, that is not the case today. Additionally, issues with funding are not uncommon, and there is less momentum behind American space exploration initiatives as compared with the 1960s [23]. Because of this, the United States needs to make some choices concerning how it maintains its interests in the field of space exploration, in order to reconcile ambitious policy statements and the reality of the capabilities of NASA.

The U.S. has recently expressed in its policy rhetoric ambivalent signals about the amount of influence it wishes to have on global space exploration. Throughout recent policy statements, the U.S. has reiterated its desire to maintain a leadership status in space, yet realities like NASA's current budget, as well as the general vagueness about the direction the U.S. space program is headed, contradict this goal. Rhetorical posturing is not a substitute for real vision and commitment. Space does not necessarily have to be a top priority in national policy, but if U.S. administrations continue to make statements about U.S. leadership in space that are reminiscent of the *Apollo* days, the American public and the rest of the world either expect the U.S. to live up to its proclamations or they will turn away, because they recognize empty promises.

Additionally, the mentality of "bigger and better for less" that has dominated the space program since the end of the *Apollo* era is harmful in the long run. Attempting to run ambitious missions that are on the same grandiose scale as before, as cheaply as possible could be detrimental not only to the quality of the American space program, but also to its reputation. If the United States is not willing to invest what is necessary into projects like a human mission to Mars, then the country needs to reevaluate its goals. Currently, it is expected that projects will run over the budgets that were originally proposed to Congress, because the ambitious goals that the United States sets out to complete are not realistic for NASA's current budget. One option is for the U.S. to take on less expensive, less ambitious projects as compared with the Apollo program. The U.S. could also partner with other nations and private companies in exchange for lower overall costs. However, if the United States desires to be the leader in space, and a pioneer in big, projects, then the country's space budget needs to be aligned with that. Either way, the program needs to be reevaluated for realistic and achievable goals.

The U.S. government and NASA should critically examine space exploration priorities and commit to implementing a program that will further realistic and robust stated policy and goals.

## 4.2 Observation 2: Struggle Between Leadership and Collaboration

Both the themes of international collaboration in space and American leadership in space have meant different things to different people over the course of the development of space exploration in the United States. They are consistently prevalent themes, but this is perhaps due to their multifaceted nature and ambiguity. International collaboration and leadership come into direct conflict in the modern era of space exploration. The end of the Cold War may have opened up the global space arena for international cooperation, but the post-Cold War era did not completely shift U.S. goals away from nationalism [24]. As the field developed and more nations became interested in participation, the United States struggled and continues to struggle to balance a desire to participate in the evolving field, but also to retain a certain amount of preeminence and control.

Although leadership surpassed competition as the new overarching goal for U.S. space exploration and leading through collaboration with other nations became a consistent goal of the U.S. space program, it is difficult to settle for being a participant rather than a leader. Blamont explains NASA's struggle between leadership and collaboration by stating that NASA generally interprets international cooperation as a situation in which the program is controlled entirely by NASA, with a few international partners on board [25]. He adds that often the international community interprets statements made by the U.S. that world problems can be solved with American leadership as a euphemism for desiring total command [25].

In the past, international collaboration in space has meant that the United States is interested in simple data sharing with other friendly, allied countries. The national goals of the United States dominated these "collaborative" efforts, and the U.S. maintained unilateral control, but yet this can still be considered as some measure of international cooperation. Earlier administrations viewed collaboration as an avenue in which the U.S. could solidify other goals like prestige and leadership, or further political interests as the atmosphere warmed between the U.S. and the U.S.S.R. Collaboration was not seen as a benefit to scientific development and the opportunity to share scientific achievement between many nations for the sake of the entire world – a more modern interpretation that resonates today.

Leadership as well has taken on many meanings over time. In the past, the United States argued for leadership in space for the sake of furthering political goals and American prestige. Now, as those specific goals are no longer relevant, NASA argues that it will "lead through cooperation." However, this sentiment is somewhat of an oxymoron. Whether or not the U.S. can successfully participate in international collaboration in space and still act as a leader in the field is entirely contingent on its definition of leadership. If leadership means that the United States must dominate the arena and maintain the upper hand in international agreements, the country will not find the true benefits to collaboration and as a result, relationships with other space-faring nations will suffer. But if the U.S. can instead find leadership to mean something more along the lines of a country that acts as a "leading participant," rather than the loudest and most important voice in the room, it may find that other nations are more willing to partner with it.

Indeed, the U.S. cannot ignore the realities of the space exploration field today, which is international and only becoming more so. To push away potential partnerships in favor of egotism would be a huge detriment to not only NASA itself, but also to world progress in space. It is important to find that balance between leader and participant. It is also important to remain open to other perspectives and allow for other nations to take the critical path on missions. True leadership, after all, is not complete domination over other parties, but rather provides the opportunity for others to participate and ensures that all voices are heard.

The U.S. should reexamine its intention to play a dominant leadership role in space exploration and consider emphasizing a commitment toward active participation in international collaboration in space.

*4.3 Observation 3: Embracing the New Paradigm*

The new paradigm of space exploration is one in which there are many international players, and in which private industry has an increasingly strong foothold. The relative role of U.S. leadership, the participation and role of other counties with interests in entering the space arena, along with privatization of space will all affect the kind of space science the U.S. conducts and in what particular aspects the U.S. will make trade-offs. Too much government control will limit competition, yet too much influence from private corporations will limit space science research and exploration to only what is commercially viable. However, it is clear that in order to be at the forefront of space exploration, the United States needs to cooperate with other prominent space-faring nations and also reduce barriers to participation in the space industry.

One area where Cold War rhetoric still dominates is in America's relationship with China in space. In this new era, China is often presented as the successor to the U.S.S.R.'s role as America's "rival" in a space race [26]. China and Russia share similarities in many ways, especially in terms of looking at how international cooperation functions in ISS-related situations [26]. Examining the historical context of Russia's inclusion on the ISS and initial U.S. reservations can aid in demonstrating how the United States could possibly extend an invitation to China [26]. It could also help evaluate whether or not the U.S. lives up to its stated goals of international collaboration. While comparing the situation with China to the past situation of the Soviet Union can be useful in understanding how national interests have shifted over time, it becomes clear that concessions will need to be made on part of the United States if it is to truly move forward in this era of globalization. Competition and rivalry in space pose a barrier to international collaboration and progress in space. With the state of current economic interdependence and a globalized world, China cannot be thought of in a simplistic, Cold War mentality [27]. In his recommendation for the U.S. and China to overcome their differences for the benefit of both parties, Hilborne states in the current global environment that is cooperative in nature, competition is not practical or logical [27].

Another area in which the U.S. needs to embrace the new paradigm is in reconsidering and possibly restructuring regulations like the International Traffic in Arms Regulations (ITAR). If international collaboration is to be the model moving forward for the United States, the presence of ITAR and similar regulations greatly hinders future successes in space exploration. Blamont explains that ITAR negatively affects collaboration by straining collaboration due to the suspicious atmosphere that the regulations cause [25]. Blamont also hints at America's suspicious nature when it comes to collaboration in space exploration, a remnant of Cold War era mentality [25]. That mentality is not keeping up with post-Cold War realities, which impedes American progress in developing positive relationships with other countries in space endeavors. This mindset also leads to potential collaborators believing it is not a good idea to place trust in the United States on long-term projects, as the U.S. presents itself as a country that puts its image and prestige before scientific successes or advancement. This is not the message we want to be conveying to the rest of the world.

While ITAR regulations are in place to protect U.S. national security and interests, some amount of deregulation does not mean dropping ITAR completely, which would render the U.S. vulnerable. Modern security concerns necessitate the need for regulations over data sharing in order ensure the safety of American intelligence and information. However, if the United States were to critically examine what particular regulations need to be in force and focus overall on improving outreach possibilities to other space-faring nations, this would demonstrate the U.S. government's actual intentions to cooperate internationally in space exploration. The reasonable way forward is to accelerate opportunities for collaboration between the United States and other interested parties, be it nations or private corporations, in order to truly capitalize on the future successes in space. Taking an active role in the private space industry and seeking out collaborative roles is essential if the U.S. space program wishes to flourish in the new paradigm.

The third arena in which it is imperative for the U.S. to fully embrace the new paradigm for space exploration is within the context of missions to Mars. If NASA is truly set on taking American astronauts to Mars, the United States should be asking realistic question about how to achieve such an ambitious collective goal. The path to Mars is undoubtedly one on which many nations must join together, simply due to the size and scope a human mission to Mars would entail. Whether the U.S. approaches missions to Mars with an international collaborative model in mind, or more like the old missions to the Moon that gave the United States a great sense of national pride, will be a true litmus test of the role the U.S. will take in these new endeavors. By looking at the framework for stated future U.S. missions to Mars, as well as what experts are recommending as courses of actions to Mars, we can see exactly how the "new paradigm" of space exploration plays out.

The United States should fully embrace the new paradigm of space exploration by - lowering barriers like ITAR that hinder the competitiveness of the American space industry, committing to collaborative endeavors with rising space-faring nations such as China and abandoning Cold War era thinking, and paving the way to Mars by encouraging the participation of many nations and space agencies on future manned missions.

## 5. Conclusion

The examination of the changing narrative of space exploration in the United States is also an examination of the changing self-perception of the country in relation to the rest of the world. Space, by the nature of the word, means thinking outside of the boundaries of our own border, be it country or global borders. Because national interests have dictated U.S. direction in space exploration, this has meant that as the country finds itself at the crux of where it stands on the global stage politically, it also does so in space endeavors.

We conclude that: competition as a driver launched the space program, but is no longer relevant with regard to Russia – and probably is not useful with regard to other countries going forward; American prestige has been a consistent but never dominant theme in space exploration rhetoric; international collaboration has been an important theme in all administrations, but it has not always had the same meaning or connotation throughout each administration; American leadership, while present in all administrations, gains momentum post-Cold War, as competition as a theme fades and the field of space exploration develops; a new paradigm of space exploration has emerged post-Cold War as an acknowledgement of changing times and more complex national and international interests.

Further study of the trends would yield additional insight. One aspect that could be investigated further is how the meaning of the five analyzed themes has changed over time. Themes like competition, prestige, collaboration, leadership, and a changing world order rarely mean exactly the same thing over the course of history. It is also often found that individual, leadership-based interpretations of the themes dictate how they are used and the value attributed to them in any given administration. As time passes, other themes may take precedence, and the approach used here could be repeated with new or additional themes in the future. It would be interesting to utilize this same method of historical narrative and content analysis to evaluate international science collaboration in other arenas, such as climate research or sustainability in the Arctic. Other areas may also yield similar situations in which the United States has transitioned from a sense of competition toward more interest in international collaboration, and scientific collaboration could be used as a tool for diplomacy.

Derived from its research, this article makes three observations to current U.S. efforts in space exploration. These observations are: 1) there is a disconnect between stated policy goals in American space exploration efforts and the implementation of those goals; 2) the United States communicates mixed messages regarding its intent to be both the dominant leader in the field of space exploration and committed as a participant in international collaboration; and 3) the United States cannot remain a true pioneer of space exploration if it does not embrace the realities of globalization and the changing dynamics within the field of space exploration.

The unknown is complicated, and space is the unknown. The United States, a country that has been at the forefront of space exploration, currently has the opportunity to look back upon the past and critically analyze history, while also analyzing the present situation, in order to determine how it should move forward into the future. This article offers three summary thoughts: 1) U.S. government and NASA should critically examine space exploration priorities and commit to implementing a program that will further realistic and robust stated policy and goals, 2) the U.S. should reexamine its intention to play a dominant leadership role in space exploration, and consider emphasizing a commitment toward active participation in international collaboration in space, and 3) the United States should fully embrace the new paradigm of space exploration by - lowering barriers like ITAR that hinder the competitiveness of the American space industry, committing to collaborative endeavors with rising space-faring nations such as China and abandoning Cold War era thinking, and paving the way to Mars by encouraging the participation of many nations and space agencies on future human missions.

## Acknowledgements

We thank Lisa Dilling for her guidance and crucial insights on the application of content analysis to the U.S. space program, as well as her review and suggestions on improving the manuscript. We also thank Scott Pace for reading the paper and making helpful suggestions to improve it. D.H. acknowledges Vicki Hunter's mentorship and guidance on her senior honors thesis project at the University of Colorado from which this paper was derived. We also thank the reviewers of this manuscript for their thoughtful and helpful suggestions to improve the text.

## References

[1]   "NASA's Vision for Space Exploration with Charles Bolden." Narrated by Neil deGrasse Tyson and Bill Nye. Interview with Charles F. Bolden, Jr. and Michael Shara. StarTalk. *StarTalk Radio Show with Neil deGrasse Tyson* (2015). http://www.startalkradio.net/show/nasas-vision-for-space-with-charles-bolden/

[2]   Alston, Giles. "Eisenhower: Leadership in Space Policy." *Reexamining the Eisenhower Presidency*, Ed. Shirley Anne Warshaw. *Greenwood* Press (1993). Print.

[3]   Kennedy, John F. "Address at Rice University on Nation's Space Effort." Rice University, Houston, Texas. *NASA JSC* (1962): 1-4. http://er.jsc.nasa.gov/seh/ricetalk.htm

[4]   Logsdon, John M. *John F. Kennedy and the Race to the Moon. Palgrave Macmillan* (2010). Print.

[5]    Ragsdale, Lyn. "Politics not Science: The U.S. Space Program in the Reagan and Bush Years." Ed. Rodger D. Launius, and Howard E. McCurdy. *Presidential Leadership and the Development of the U.S. Space Program. NASA History Office* (1994). Print.

[6]    Logsdon, John M. "Space in the Post-Cold War Environment," *NASA History* (1992): 89-102. http://history.nasa.gov/sp4801-chapter6.pdf

[7]    Hsieng, Hsiu-Fang, and Sarah E. Shannon. *Three Approaches to Qualitative Content Analysis. Qualitative Health Research.* (2005): 1277-1288. http://journals.sagepub.com/doi/pdf/10.1177/1049732305276687

[8]    Johnson, Lyndon B. "The Vice President Answers." *NASA History* (1961): 1-4. http://www.au.af.mil/au/awc/awcgate/key_docu.htm

[9]    National Security Council. "Implications of the Soviet Earth Satellite for U.S. Security (NSC 5520)." *George C. Marshall Institute* (1955): 1-20. http://marshall.wpengine.com/wp-content/uploads/2013/09/NSC-5520-Statement-of-Policy-on-U.S.-Scientific-Satellite-Program-20-May-1955.pdf

[10]   Nixon, Richard. "Statement About the Future of the United States Space Program," *The American Presidency Project* (1970): 1-4. http://www.presidency.ucsb.edu/ws/?pid=2903

[11]   Obama, Barack. "Remarks by the President on Space Exploration in the 21st Century." John F. Kennedy Space Center, Merritt Island. *Whitehouse.gov.* (2010): 1-6. https://www.whitehouse.gov/the-press-office/remarks-president-space-exploration-21st-century

[12]   Brookings Institute, The. "Proposed Studies on the Implications of Peaceful Space Activities for Human Affairs." *ProQuest Congressional* (1961): 1-284. http://congressional.proquest.com/congressional/docview/t21.d22.cmp-1961-sah-0002?accountid=14503

[13]   Sheldon, Charles S. II. "United States and Soviet Rivalry in Space: Who Is Ahead, and How Do the Contenders Compare." *Congressional Research Service Reports* (1969): 1-54. http://congressional.proquest.com/congressional/docview/t21.d22.crs-1969-spx-0004?accountid=14503

[14]   National Security Council. "U.S. Policy on Outer Space (NSC 5918)." *George C. Marshall Institute* (1959): 1-26. http://marshall.wpengine.com/wp-content/uploads/2013/09/NSC-5918-1-U.S.-Policy-on-Outer-Space-26-Jan-1960.pdf

[15]   Bush, George H. W. "Remarks on the 20th Anniversary of the Apollo 11 Moon Landing." *The American Presidency Project* (1989): 1-4. http://www.presidency.ucsb.edu/ws/?pid=17321

[16]   National Security Action Memorandum. "U.S.-U.S.S.R. Cooperation in the Exploration of Space (NSAM 129)." *Federation of American Scientists* (1962): 1. http://fas.org/irp/offdocs/nsam-jfk/nsam-129.htm

[17]   Nixon, Richard. "Statement About the Future of the United States Space Program," *The American Presidency Project* (1970): 1-4. http://www.presidency.ucsb.edu/ws/?pid=2903

[18]   Reagan, Ronald. "Remarks at the 25th Anniversary Celebration of the National Aeronautics and Space Administration." *The American Presidency Project* (1983): 1-5. http://www.presidency.ucsb.edu/ws/index.php?pid=40662

[19]   National Space Council. "National Space Policy Directives and Executive Charter (NSPD-1)." *NASA History* (1989): 1-14. http://fas.org/spp/military/docops/national/nspd1.htm

[20]   National Commission on Space. "Pioneering the Space Frontier: An Exciting Vision of Our Next Fifty Years in Space." *NASA History* (1986): 1-225. https://www.nasa.gov/pdf/383341main_60%20-%2020090814.5.The%20Report%20of%20the%20National%20Commission%20on%20Space.pdf

[21]   Obama, Barack. "Remarks by the President on Space Exploration in the 21st Century." John F. Kennedy Space Center, Merritt Island. *Whitehouse.gov.* (2010): 1-6. https://www.whitehouse.gov/the-press-office/remarks-president-space-exploration-21st- century

[22]   111th United States Congress. "National Aeronautics and Space Administration Authorization Act of 2010." *NASA History* (2010): 1-42. http://www.nasa.gov/pdf/649377main_PL_111-267.pdf

[23]   Dinerman, Taylor. "Space policy versus space politics: lessons for the future." *The Space Review* (2010): 1-3. http://www.thespacereview.com/article/1568/1

[24]   Logsdon, John M. "Space in the Post-Cold War Environment," *NASA History* (1992): 89-102. http://history.nasa.gov/sp4801-chapter6.pdf

[25]   Blamont, Jacques. "International Space Exploration: Cooperative or Competitive?" *Space Policy* (2005): 89-92. 10.1016/j.spacepol.2005.03.003

[26]   Zhao, Yun. "Legal Issues of China's Possible Participation in the International Space Station: Comparing to the Russian Experience." Journal of East Asia & International law 6.1 (2013): 155-174. http://web.b.ebscohost.com/ehost/pdfviewer/pdfviewer?sid=5f4c9d62-7d4e-4284-9fb6-870f73c49a78%40sessionmgr111&vid=0&hid=124

[27]   Hilborne, Mark. "China's Rise in Space and U.S. Policy Responses: A Collision Course?" *Space Policy* (2013): 121-127. 10.1016/j.spacepol.2013.03.005

*APPENDIX: DOCUMENTS ANALYZED IN CONTENT ANALYSIS*

| Administration | Author | Title | Year | Document Category |
|---|---|---|---|---|
| Eisenhower | National Security Council | Implications of the Soviet Earth Satellite for U.S. Security (NSC 5520) | 1955 | Policy Recommendation |
| Eisenhower | Dwight D. Eisenhower | Official White House Transcript of President Eisenhower's Press and Radio Conference #123 | 1957 | Leadership Statement |
| Eisenhower | Dwight D. Eisenhower | 1958 State of the Union Address | 1958 | Leadership Statement |
| Eisenhower | 85th United States Congress | National Aeronautics and Space Act of 1958 | 1958 | Policy |
| Eisenhower | National Security Council | U.S. Policy on Outer Space (NSC 5918) | 1959 | Policy |
| Kennedy | The Brookings Institute | Proposed Studies on Implications of Peaceful Space Activities for Human Affairs | 1961 | Policy Recommendation |
| Kennedy | Lyndon B. Johnson | Policy Recommendation to President Kennedy | 1961 | Policy Recommendation |
| Kennedy | John F. Kennedy | Special Message to the Congress on Urgent National Needs | 1961 | Leadership Statement |
| Kennedy | National Security Action Memorandum | U.S.-U.S.S.R. Cooperation in the Exploration of Space (NSAM 129) | 1962 | Policy |
| Kennedy | John F. Kennedy | Address at Rice University on Nation's Space Effort | 1962 | Leadership Statement |
| Nixon | Charles S. Sheldon II | United States and Soviet Rivalry in Space: Who Is Ahead, and How Do the Contenders Compare | 1969 | Policy Recommendation |
| Nixon | Space Task Group | Report of the Space Task Group | 1969 | Policy Recommendation |
| Nixon | Richard Nixon | Statement About the Future of the United States Space Program | 1970 | Leadership Statement |
| Nixon | National Security Decision Memorandum | Exchange of Technical Data between the United States and the International Space Community (NDSM 72) | 1970 | Policy |
| Nixon | Richard Nixon | Announcement on the Space Shuttle | 1972 | Leadership Statement |
| Reagan | Ronald Reagan | Remarks at the 25th Anniversary Celebration of the National Aeronautics and Space Administration | 1983 | Leadership Statement |
| Reagan | Ronald Reagan | 1984 State of the Union Address | 1984 | Leadership |

| | | | | Statement |
|---|---|---|---|---|
| Reagan | National Commission on Space | Pioneering the Space Frontier | 1986 | Policy Recommendation |
| Reagan | Sally K. Ride | Leadership and America's Future in Space | 1987 | Policy Recommendation |
| Reagan | Ronald Reagan | Presidential Directive on National Space Policy | 1988 | Policy |
| Bush Sr. | George H. W. Bush | Remarks on the 20th Anniversary of the Apollo 11 Moon Landing | 1989 | Leadership Statement |
| Bush Sr. | George H. W. Bush | Proclamation 5999 - Space Exploration Day | 1989 | Leadership Statement |
| Bush Sr. | National Space Council | National Space Policy Directives and Executive Charter (NSPD-1) | 1989 | Policy |
| Bush Sr. | U.S. Congress Office of Technology | Exploring the Moon and Mars: Choices for the Nation | 1991 | Policy Recommendation |
| Bush Sr. | Vice President's Space Policy Advisory Board | A Post Cold War Assessment of U.S. Space Policy | 1992 | Policy Recommendation |
| Clinton | Bill Clinton | Statement on the Space Station Program | 1993 | Leadership Statement |
| Clinton | Bill Clinton | Proclamation 6707 - National Apollo Anniversary Observance | 1994 | Leadership Statement |
| Clinton | The White House National Science and Technology Council | Fact Sheet: National Space Policy | 1996 | Policy |
| Clinton | David P. Radzanowski and Marcia S. Smith | Clinton and Bush Administration National Space Policies: A Comparative Analysis | 1997 | Policy Recommendation |
| Clinton | 106th U.S. Congress | 2000 NASA Authorization Act | 2000 | Policy |
| Bush Jr. | George W. Bush | Remarks on U.S. Space Policy | 2004 | Leadership Statement |
| Bush Jr. | President's Commission on Implementation of United States Space Exploration Policy | A Journey to Inspire, Innovate, and Discover | 2004 | Policy Recommendation |
| Bush Jr. | National Aeronautics and Space Administration | The Vision for Space Exploration | 2004 | Policy |
| Bush Jr. | George W. Bush | U.S. National Space Policy | 2006 | Policy |

| Bush Jr. | Deborah D. Stine | U.S. Civilian Space Policy Priorities: Reflections 50 Years After Sputnik | 2007 | Policy Recommendation |
|---|---|---|---|---|
| Obama | U.S. Human Spaceflight Plans Committee | Seeking a Human Spaceflight Program Worthy of a Great Nation | 2009 | Policy Recommendation |
| Obama | Barack Obama | Remarks By the President on Space Exploration in the 21st Century | 2010 | Leadership Statement |
| Obama | Barack Obama | U.S. National Space Policy | 2010 | Policy |
| Obama | 111th Congress | NASA Authorization Act | 2010 | Policy |
| Obama | Charles Bolden | State of NASA | 2015 | Leadership Statement |